\documentclass[useAMS,usenatbib]{mn2e}

\def \be{\begin{equation}}
\def \ee{\end{equation}}
\def \ben{\begin{eqnarray}}
\def \een{\end{eqnarray}}

\title{Evolution of primordial black holes in Jordan-Brans-Dicke cosmology}
 
\author[A. S. Majumdar, D. Gangopadhyay and L. P. Singh]{A. S. Majumdar$^1$\thanks{Email: archan@bose.res.in}, D. Gangopadhyay$^1$ and L. P. Singh$^2$\\
$^1$S. N. Bose National Centre for Basic Sciences,
Salt Lake, Kolkata 700 098, India\\
$^2$Physics Department, Utkal University, Vani Vihar, Bhubaneswar 
751004, India}

\begin{document}
 
\maketitle

\label{firstpage}

\begin{abstract}
We consider the evolution of primordial black holes in a generalized 
Jordan-Brans-Dicke cosmological model where both the Brans-Dicke
scalar field and its coupling to gravity are dynamical functions determined 
from the evolution equations. The evaporation rate for the black holes is 
different from the case in standard cosmology. We show that the accretion of 
radiation can proceed effectively in the radiation dominated era. It follows
that the black hole lifetime shortens for low initial mass, but increases 
considerably for larger initial mass, thus providing a mechanism for the 
survival of primordial black holes as candidates of dark matter. We derive a 
cut-off value for the initial black hole mass, below which primordial black 
holes evaporate out in the radiation dominated era, and above which they 
survive beyond the present era. 
\end{abstract}


\section{Introduction}

The Jordan-Brans-Dicke (JBD) \citep{brans} theory is one of the earliest and 
most
well-motivated alternatives to the Einstein theory of gravitation. The value
of the gravitational ``constant'' is set by the inverse of a time-dependent
classical scalar field with a coupling parameter $\omega$. General 
relativity is recovered in the limit of $\omega \to \infty$. Solar system
observations impose lower bounds on $\omega$ \citep{obser1,obser}. 
Generalized JBD models are obtained in the low energy limit of higher 
dimensional theories. String theoretic \citep{string} and 
Kaluza-Klein \citep{kk} models
after compactification of the extra dimensions yield several variants of
JBD models (or general scalar-tensor models) in which the the scalar field
coupling $\omega$ may become dynamical,
and also models with potential terms for the JBD scalar field.

JBD models have been used for tackling several cosmological
problems pertaining to different eras of evolution of the 
Friedman-Robertson-Walker (FRW) universe. The scenario of extended
inflation \citep{extended1,extended2} was proposed within the context 
of JBD cosmology. 
Generalized JBD models resulting from the compactification of
higher dimensional actions \citep{asm11,asm12,asm13} enable 
efficient transition to the
post-inflationary radiation dominated phase (or reheating).
The JBD scalar field has also been incorporated in quintessence 
scenarios for obtaining the present acceleration of the 
universe \citep{quint1,quint2,quint3}. A more general class of JBD theory where 
the coupling parameter $\omega$ is an arbitrary function of the scalar 
field \citep{bergmann1,bergmann2,bergmann3} has interesting consequences 
on cosmic 
evolution \citep{sahoo1,sahoo2} and has also been applied in the context 
of obtaining
singularity free cosmology \citep{kalyana} and relic gravitational 
waves \citep{barrow,sahoo3}. 

Primordial black holes (PBHs) are potentially fascinating 
cosmo-archeological tools \citep{pbh}. It is possible for  
PBHs to impact through
their evaporation products diverse cosmological 
processes 
such as  baryogenesis and 
nucleosynthesis \citep{baryo1,baryo2,baryo3,baryo7,baryo4,baryo5,baryo6},  the 
cosmic microwave background
radiation \citep{cmbr1,cmbr2}, and  the
growth of perturbations as well \citep{afshordi}. PBHs could act as 
seeds for structure 
formation \citep{ostriker} and could also form a 
significant component of 
dark matter through efficient early accretion in braneworld 
scenarios \citep{brane1,brane2,brane3,brane4}.  Mechanisms for growth of
supermassive black holes by PBHs accreting dark energy have been 
proposed \citep{bean}, though recent results by 
\citet{harada11} and \citet{harada12} indicate the lack of self-similar growth 
of black holes by
accreting quintessence. 

The feasibility of black hole solutions
in JBD theory was first discussed by \citet{hawking}.
The coexistence
of black holes with a long range scalar field in cosmology has
interesting consequences such as the possibility of energy exchange between
the black holes and the scalar field \citep{bhscalar}. 
The black holes themselves
could be used to constrain the variation of fundamental constants \citep{mac}.  
Another interesting 
issue of gravitational memory of black holes in JBD theory has also
been studied \citep{harada2}. The
evolution of PBHs in JBD cosmology has however remained unexplored 
in the literature.
Since JBD models entail modification of
the standard cosmological evolution, it is expected that the evolution
of PBHs in JBD models and their associated consequences on various 
cosmological processes could be substantially modified compared to the
case of standard cosmology.  

The purpose of this work is to perform a preliminary study of
the evolution of PBHs in JBD cosmology.   
We consider a generalized scalar 
tensor theory \citep{bergmann1,bergmann2,bergmann3}
where the coupling parameter $\omega(\phi)$ is a function of the scalar
field $\phi$, and is thus also time-dependent. We use a
set of power-law solutions for the scale factor, the JBD field and the coupling
$\omega$ obtained in the radiation era \citep{sahoo1,sahoo2}, and 
investigate the net 
energy flux for the PBHs resulting from the processes of Hawking evaporation
and the acretion of radiation. We show that accretion of radiation can
be effective in the radiation dominated era since the growth of PBHs 
is always subdominant to the growth of the cosmological horizon
in this JBD model. In the matter 
dominated era, the dynamics of the PBHs is solely
governed by the evaporation process whose rate is determined by another 
set of power-law solutions of this JBD model in the matter dominated 
era \citep{sahoo1,sahoo2}.
The lifetime for PBHs are obtained
in terms of their initial masses. Our results outline a viable scenario for
the growth and survival of PBHs as constituents of dark matter in JBD 
cosmology.   

\section{Generalized Brans-Dicke model}

We begin with a generalized JBD action given by
\ben
S = \frac{1}{16\pi}\int d^4x \sqrt{-g}[\phi R - \frac{\omega(\phi)}{\phi}
(\partial_{\mu}\phi)^2] + S_{\mathrm{matter}}
\label{action}
\een 
where the strength of the running gravitational coupling is given by the 
inverse of
$\phi$, i.e., at any time $t$, $G = 1/\phi(t)$, and $S_{\mathrm{matter}}$ corresponds to the action of the relativistic
fluid of particles ($p=\rho/3$) in the flat ($k=0$) FRW early universe with 
scale factor $a$. The Friedman equation and the equation of motion for
the JBD field $\phi$ obtained from the above action are given respectively by
\ben
\frac{\dot{a}^2}{a^2} + \frac{\dot{a}\dot{\phi}}{a\phi} - 
\frac{\omega \dot{\phi}^2}{6\phi^2} = \frac{\rho}{3\phi}
\label{fe}
\een
and
\ben
\ddot{\phi} + 3\frac{\dot{a}\dot{\phi}}{a} = \frac{\rho -3p}{2\omega +3} -\frac{\dot{\omega}\dot{\phi}}{2\omega + 3}
\label{eqmotion}
\een
with the energy conservation equation given by
\be
\dot{\rho} +3\frac{\dot{a}}{a}(\rho + p) =0
\label{enconv}
\ee 
For a radiation dominated evolution, one has $\rho \approx \rho_R \propto a^{-4}$, and $p = \rho/3$. Assuming power law solutions for the scale factor $a(t)$,
the JBD field $\phi(t)$, and also the coupling parameter $\omega(t)$ 
one can obtain \citep{sahoo1,sahoo2} the following time-dependences 
of these quantities:
\be
a(t) = a_i\biggl(\frac{t}{t_i}\biggr)^{\frac{3}{\omega_i +6}}
\label{rdsoln1}
\ee
\be
\phi(t) = \phi_i\biggl(\frac{t}{t_i}\biggr)^{-\frac{3}{\omega_i +6}}
\label{rdsoln2}
\ee
\be
2\omega(t)+ 3 = (2\omega_i +3)\biggl(\frac{t}{t_i}\biggr)^{\frac{2\omega_i + 3}{\omega_i +6}}
\label{rdsoln3}
\ee
where the subscript $i$ indicates the initial values of the variables. In order
to minimize the departure of the evolution of the scale factor from the case
of the radiation dominated evolution of standard 
cosmology (i.e., $a(t) \propto t^{1/2}$)
we choose $|\omega_i| \ll 1$. Without loss of generality, we set
$\phi_i = \xi M_{pl}^{2}$, where $M_{pl} = G^{-1/2}$ is the Planck mass
(corresponding to the present value of $G$). 
Note that the solution for the JBD field $\phi$ indicates the strengthening of
gravity with the increase of cosmic time. Such a behaviour for $\phi$ arises
not just from the specific model (\ref{action}) used here, but is a rather
generic feature of several generalized JBD or scalar tensor models with
a potential for the scalar field which rolls down the slope leading to
the increase of $G$ with time \citep{asm11,asm12,asm13}.

\section{Evolution of primordial black holes}

Primordial black holes could form due to various mechanisms \citep{formation}
some of which have been studied in detail in 
the literature \citep{zeldovich1,zeldovich2,zeldovich3,zeldovich4,zeldovich5}.
The gravitational collapse of perturbations within the cosmological horizon
could lead to sub-horizon sized black hole formation after Jeans crossing.
Such a mechanism may not be very effective however, in JBD cosmology where
the Jeans length itself could be significantly increased due to the weakening
of gravity in certain models by a large value of the JBD field in the early 
universe. Nevertheless, a strong (first order) inflationary phase transition
in extended inflation models driven by the JBD 
field \citep{extended2,asm11,asm12,asm13} naturally leads to  
unnucleated and trapped false vacuum regions and 
topological defects such as domain walls and wormholes
that could easily collapse into black holes, and could also create
super-horizon density 
perturbations. The formation of super-horizon
scale PBHs in the expanding FRW background has been studied 
recently \citep{harada3}.
In the present analysis we will not take 
recourse to any particular
formation mechanism for PBHs, but 
rather study their cosmological evolution
in JBD theory, assuming that there exist PBHs in such scenarios.  

We now consider the evolution of PBHs in the cosmological background governed
by the above solutions (\ref{rdsoln1}),(\ref{rdsoln2}) and (\ref{rdsoln3}). 
We assume that the PBH density is low enough to ensure radiation domination.
For a PBH immersed in the
radiation field, the 
accretion of radiation leads to the increase of its mass with the rate
given by
\be
\dot{M}_{acc} = 4\pi f r_{BH}^2\rho_R
\label{accr1}
\ee
where $r_{BH}=2M/\phi$ is the black hole radius, and $f$ is the accretion
efficiency. Using the solution for $\phi$ given by Eq.(\ref{rdsoln2}) with
the assumption of $|\omega_i| \ll 1$, and using 
$\rho_R = [(3M_{pl}^2)/(32\pi t^2)]$, one obtains
\be
\dot{M}_{acc} = \frac{BM^2}{t}
\label{accr2}
\ee
where $B=(3f)/(2\xi^2M_{pl}^2t_i)$. Eq.(\ref{accr2}) is integrated
to yield
\be
\frac{M}{M_i} = \frac{1}{1-BM_i \mathrm{ln}(t/t_i)}
\label{accr3}
\ee
with $M_i$ being the initial mass of the PBH at time $t_i$.
Since the logarithmic growth rate for a PBH given by Eq.(\ref{accr3}) 
is subdominant to the linear growth of the horizon mass $M_H \sim t$ 
(since $a \sim t^{1/2}$), once a PBH is formed, it
is indeed possible for it to grow in size by accreting the radiation energy
within its cosmological horizon.  

For a complete picture of PBH evolution, one needs also to consider the
Hawking evaporation process, whose rate is given by
\be
\dot{M}_{evap} = - 4\pi g r_{BH}^2 \sigma T^4
\label{evap1}
\ee
where $T=\phi/(8\pi M)$ is the Hawking temperature, $\sigma$ the 
Stefan-Boltzmann
constant and $g$ is the effective number of degrees of freedom of the particles
emitted by the black hole. The solution for $\phi$ given 
 by Eq.(\ref{rdsoln2}) leads to
\be
\dot{M}_{evap} = -\frac{A}{M^2t}
\label{evap2}
\ee
where $A = (g \sigma \xi^2 M_{pl}^4 t_i)/(256\pi^3)$. The complete 
evolution for 
the PBH
is thus described by the combination of Eqs.(\ref{accr2}) and (\ref{evap2}):
\be
\dot{M} = -\frac{A}{M^2t} + \frac{BM^2}{t}
\label{bheq}
\ee
It is apparent from Eq.(\ref{bheq}) that for PBHs with initial mass
$M_i << (A/B)^{1/4}$, the rate of evaporation exceeds that of accretion.
For such a case, accretion soon becomes negligible
and black holes lose energy at a rate given effectively by Eq.(\ref{evap2}). 
Note that though the rate of evaporation decreases with time, it is
still higher than the corresponding rate in standard cosmology. This
is because the Hawking temperature $T=\phi/(8\pi M)$ is larger for
JBD PBHs for large $\phi$. Hence, a PBH with initial mass $M_i < (A/B)^{1/4}$
evaporates out much quicker, with a lifetime $t_{evap}$ given by
\be
\frac{t_{evap}}{t_i} = \mathrm{exp}\biggl[\frac{1}{3A}\biggl(\frac{M_i}{M_{pl}}\biggr)^3\biggl(\frac{t_{pl}}{t_i}\biggr)\biggr]
\label{lifetime1}
\ee
However, PBHs with initial mass $M_i > (A/B)^{1/4}$ experience monotonic 
growth with accretion dominating over
evaporation throughout the period of validity of Eq.(\ref{bheq}), i.e., 
throughout the radiation dominated era. Eq.(\ref{bheq}) can be integrated 
exactly and leads to the
following mass-time relationship for the PBHs:
\ben
\mathrm{ln}(\frac{t}{t_i}) &=& \frac{1}{2BC^{1/4}}\biggl[\mathrm{tan}^{-1}
\biggl(\frac{M}{C^{1/4}}\biggr) - \mathrm{tan}^{-1}\biggl(\frac{M_i}{C^{1/4}}\biggr)\biggr]\nonumber\\
&+& \frac{1}{4BC^{1/4}}\mathrm{ln}\vert\frac{(M-C^{1/4})(M_i+C^{1/4})}{(M_i-C^{1/4})(M+C^{1/4})}\vert
\label{bhsol1}
\een
with $C=A/B$. Accretion of radiation can proceed 
effectively till the universe stays 
radiation dominated, i.e., up to the era of matter radiation equality $t_{eq}$.
This result is qualitatively different from the widely accepted picture
in the standard cosmological evolution where accretion of radiation in 
the radiation dominated era seems to be ineffective \citep{pbh}. The
domination of accretion over evaporation is observed also in other 
modified gravity
theories, such as in the braneworld 
scenario \citep{brane1,brane2,brane3,brane4}.
The maximum mass achieved by a PBH of initial mass $M_i$ is given by
\be
\frac{M_{max}}{M_i} \approx \frac{1}{1-BM_i{\mathrm ln}(t_{eq}/t_i)}
\label{maxmass}
\ee

In the matter dominated era, $\rho_M \sim a^{-3}$ and $p=0$. A set of 
solutions for the JBD cosmological equations  (\ref{fe}),
(\ref{eqmotion}) and (\ref{enconv})  is given by \citep{sahoo1,sahoo2}
\be
a(t) = a(t_{eq})\biggl(\frac{t}{t_{eq}}\biggr)^{\frac{2}{3}}
\label{mdsoln1}
\ee
\be
\phi(t) = \phi(t_{eq})\biggl(\frac{t}{t_{eq}}\biggr)^{-\frac{4}{3}}
\label{mdsoln2}
\ee
\be
\omega(t) = -\omega(t_{eq})\biggl(\frac{t}{t_{eq}}\biggr)^{\frac{4}{3}}
\label{mdsoln3}
\ee
Matching these solutions with those obtained in the radiation dominated
era given by Eqs.(\ref{rdsoln1}),(\ref{rdsoln2}) and (\ref{rdsoln3}), 
one has $\phi(t_{eq}) = \xi M_{pl}^2(t_i/t_{eq})^{1/2}$ and 
$\omega(t_{eq}) = \omega_i(t_{eq}/t_i)^{1/2}$. Substituting the solution for
the JBD field $\phi$ in Eq.(\ref{evap1}) leads to the following evaporation
law for PBHs in the matter dominated era:
\be
\dot{M} = -\tilde{A}\frac{1}{M^2 t^{8/3}}
\label{mdevap1}
\ee
with $\tilde{A} = A t_i M_{pl}^4 t_{eq}^{5/3}$. The mass of the PBHs  
evolve as
\be
M(t) = M_{max}\biggl[1- \frac{3\tilde{A}}{5M^3_{max}}\biggl(t_{eq}^{-5/3}-t^{-5/3}\biggr)\biggr]^{1/3}
\label{bhsolmd}
\ee
in the matter dominated era. The lifetime for the PBHs $t_{evap}$ is given by
\be
t_{evap} = t_{eq}\biggl[1- \frac{5}{3}\frac{M^3_{max}}{\tilde{A}t^{-5/3}_{eq}}\biggr]^{-3/5}
\label{lifetime20}
\ee
which upon (using Eq.(\ref{maxmass})) reduces to
\be
t_{evap} = t_{eq}\biggl[1- \frac{5}{3}\frac{M_i^3}{\bigl(1-BM_i{\mathrm ln}(t_{eq}/t_i)\bigr)^3\bigl(At_iM_{pl}^4\bigr)}\biggr]^{-3/5}
\label{lifetime2}
\ee
Since the availability of background radiation diminishes substantially
in the matter
dominated era, one can safely neglect any accretion term corresponding to
the accretion of radiation by the PBHs. However, other forms of accretion
such as of the energy of a quintessence field could, in principle, be
effective \citep{bean}, that we do not consider in the present analysis. 
The rate of evaporation is again faster, as in the radiation dominated era,
than in the case of standard cosmology (where $ t_{evap} \sim M^3_{max}$).  
However, the rate decreases with time
as is evident from the r.h.s of 
Eq.(\ref{mdevap1}) whose counterpart in the standard theory is independent
of time. 

The overall lifetime of  a JBD PBH has an interesting comparison to a PBH
lifetime in standard cosmology. A JBD PBH with initial 
mass $M_i < (A/B)^{1/4}$,
evaporates out quicker than a PBH in standard cosmology, being 
unable to accrete in the radiation dominated era. However, when the initial
mass exceeds $(A/B)^{1/4}$, accretion proceeds effectively and
dominates over evaporation in 
the radiation dominated era
increasing the PBH mass at the time of matter-radiation equality. 
In the matter dominated era, though the evaporation rate of JBD PBH  is faster,
it starts to evaporate out much later than in standard cosmology 
(without accretion). Thus, the lifetime for a PBH
with $M_i > (A/B)^{1/4}$ is enhanced in the JBD scenario compared to
the case of standard cosmology.
Note that in the present analysis we have not considered any 
possible back reaction of the PBHs on the local background value of the
JBD field resulting from the local change of energy density $\rho$ due to the 
PBHs. The form of the JBD field equation (\ref{eqmotion}) is such that
the $\rho$-dependence drops out in the radiation dominated era. 
Further, in the
late matter-dominated era the value of $\omega$ is large enough (in order to
satisfy present observational bounds) such that the $\rho$-dependent 
term again becomes
negligible in Eq.(\ref{eqmotion}). So any back reaction could be 
effective only in the
early matter dominated epoch during which
a local rise in energy density could result in a slower rate of decrease 
for the
$\phi$ field near the PBHs than what is obtained 
in Eq.(\ref{mdsoln2}). The resultant change in the PBH evaporation law 
(\ref{mdevap1}) could introduce a minor correction to the PBH lifetime 
(\ref{lifetime2}) that we have neglected here.

It would be interesting to observe particular examples of evolving
PBHs in JBD cosmology. Note that the accretion and evaporation
rates depend on the initial value $\phi$ of the JBD field  through
the quantities $A$ and $B$ defined earlier. Here we fix the value of $\phi_i$
or $\xi$ by setting the present value of $\phi(t_{now}) = M^2_{pl}$ 
(corresponding
to the present strength of gravity) and evolving $\phi$ given
by Eqs.(\ref{mdsoln2}) and (\ref{rdsoln2}) corresponding to the matter
and radiation dominated eras, respectively, backwards in time. We thus
obtain the value of $\phi(t_{EW}) = 10^{18}M^2_{pl}$, where $t_{EW}$ denotes
the electroweak era since which the JBD evolution
equations (\ref{eqmotion}) are assumed to be valid. From
Eq.(\ref{lifetime1}) it follows that for a PBH with $M_i < (A/B)^{1/4}$
to survive the radiation dominated era, $M_i \ge 10^{13}M_{pl}$. A PBH which
forms at time $t_{EW}$ with an initial mass of the order of the cosmological
horizon mass at $t_{EW}$ satisfies the condition $M_i > (A/B)^{1/4}$, and
accretes radiation. Though the actual growth of mass turns out to be
negligible for such a PBH,  accretion  does play an important role enabling
it to survive for much longer ($t_{evap} > t_{now}$). 

Another interesting
consequence of Eq.(\ref{lifetime2}) is that for a PBH to
evaporate during the present era, i.e., $t_{evap} = t_{now}$ one needs
to have $M_{max} \approx 10^{10}M_{pl}$. Since accretion cannot cause
significant growth, one requires $M_i \sim M_{max} \approx 10^{10}M_{pl}$.
However, a PBH with such a small initial mass cannot even survive the
radiation dominated era. Only those PBHs which have $M_i > 10^{13}M_{pl}$
enter the matter dominated era and these also survive up to the present
era, i.e., $t_{evap} >> t_{now}$. Thus there are no end products
of black hole evaporation to contend with in JBD cosmology during the
matter dominated era and beyond up to present times. All PBHs with initial
mass $M_i < 10^{13}M_{pl}$ evaporate out in the radiation dominated era
itself. On the other hand, PBHs with larger mass can survive as intermediate
mass astrophysical black holes contributing to the dark matter density.     

\section{Conclusions}

To summarize, in this letter we have studied the evolution of PBHs in
JBD cosmology and obtained a cut-off value for the initial mass which
decides whether the PBHs could survive up to 
today. An interesting consequence of
our analysis is that no PBHs evaporate during the matter dominated era,
and hence there is no distortion of the cosmic microwave background 
radiation spectrum due
to evaporating PBHs. Though our results are in the 
context of
a particular JBD model, the sort of evolution obtained for the JBD field
$\phi$ is generic to other JBD models as well, in particular those 
following from higher dimensional theories \citep{asm11,asm12,asm13}.
In order to satisfy 
solar system constraints \citep{obser} such as the one recently obtained
through the frequency shift of radio photons to and from the Cassini
spacecraft as they passed near the sun \citep{obser1},
one requires $|\omega(t_{now})| > 10^4$, which using
Eqs. (\ref{mdsoln3}) and
(\ref{rdsoln3}), leads to
$\omega_i \sim 10^{-14}$,  thus strongly validating our approximation
of $|\omega_i| \ll 1$ used to derive the black hole evolution equation
in the radiation dominated era. With the possibility of future observations 
of actually finding or ruling out JBD models by determining 
$\omega$ \citep{obser2}, it would indeed be exciting to re-work
the standard observational 
constraints \citep{pbh,baryo6,formation} on the density of
PBHs at several cosmological eras 
in the JBD scenario. In particular, the faster evaporation of smaller PBHs
in the radiation dominated era could impose tighter constraints on the initial
mass spectrum from nucleosynthesis bounds \citep{baryo6}.
Investigation of these issues calls for a more comprehensive analysis by 
considering a population 
of black holes
in the early universe evolving according to the JBD dynamics.

\label{lastpage}
 
\end{document}